\documentstyle[prd,aps,twocolumn,floats]{revtex}
\renewcommand{\epsilon}{\varepsilon}
\begin{document}
\twocolumn[\hsize\textwidth\columnwidth\hsize\csname
@twocolumnfalse\endcsname
\preprint{}
\title{Covariant cosmological perturbation dynamics in the  
large-scale limit}
\author{Winfried Zimdahl}
\address{Fakult\"at f\"ur Physik, Universit\"at Konstanz, PF 5560 M678
D-78457 Konstanz, Germany}
\author{Diego Pav\'{o}n}
\address{Departamento de F\'{\i}sica,
Universitat Aut\`{o}noma de Barcelona,
E-08193 Bellaterra (Barcelona), Spain}
\date{\today}
\maketitle
\begin{abstract}

Using the existence of a covariant conserved quantity on large  
perturbation scales in a spatially flat perfect fluid or scalar  
field universe, we present a general formula for gauge-invariantly  
defined comoving energy density perturbations which encodes the  
entire linear perturbation dynamics in a closed time integral.
On this basis we discuss perturbation modes in different  
cosmological epochs.
\end{abstract}
\vspace{1.5cm}
]

\section{Introduction}
Because of its notable significance for the early stages of structure  
formation in the universe cosmological perturbation theory has  
received considerable attention during the past decades (see, e.g.  
\cite{KoSa,MFB,Ellis,LL}).

Depending on the choice of basic perturbation variables there are  
different possibilities to formulate the dynamics of primordial
inhomogeneities.
These choices may refer to different gauges or to different sets of  
gauge-invariant variables.
In this paper we focus on the covariant approach which uses exactly  
defined tensorial quantities
(\cite{Ols,WoKu,EB,Jack}) instead of conventional, generally  
gauge-dependend perturbation variables.
A covariant approach is conceptionally superior to noncovariant
approaches  since it avoids the explicit introduction of a fictitious  
background universe and thus circumvents the gauge-problem, which  
just originates from the nonuniqueness of the conventional splitting  
of the spacetime into a homogeneous and isotropic zeroth order and  
first-order perturbations about this background.
The basic dynamic quantity used by Olson \cite{Ols}, Woszczyna and  
Kulak \cite{WoKu}, Ellis and Bruni \cite{EB} and Jackson \cite{Jack}  
is the covariantly defined spatial gradient of the energy density  
of a comoving (with the four-velocity of the cosmic fluid) observer.  
The corresponding dimensionless, fractional quantity obeys a  
second-order differential equation (see, e.g. \cite{Jack}).

As was shown recently \cite{ZCQG}, the linear cosmological  
perturbation theory of an almost homogeneous and isotropic universe  
may be simplified by introducing covariant variables defined with  
respect to hypersurfaces of constant expansion, constant curvature  
or constant energy density.
These new quantities which represent suitable linear combinations  
of Ellis-Bruni-Jackson type variables turned out to be particularly  
useful to characterize a conserved quantity on large perturbation  
scales which was
applied subsequently
to study the perturbation dynamics in an inflationary universe.  
\cite{ZPRD}.
Generally, the importance of conserved quantities is well  
recognized in the literature
(\cite{Bardeen,BST,Ly,H1,H2,DB}).
In this paper we clarify how the existence of a covariant conserved   
quantity may be used to solve the perturbation equations in the  
corresponding limit.
Our main purpose is to derive a general formula which reduces the  
entire large-scale linear perturbation dynamics for arbitrary  
equations of state to a closed time integral over homogeneous  
``background'' quantities.
We then demonstrate how the cosmological mode structure for all  
cases of interest (exponential inflation, power-law inflation,  
radiation and matter dominated periods etc.) follow from the general  
formula in an elementary way.

This paper is organized as follows.
Section II briefly recalls the basic relations of a covariant  
perturbation theory with special emphasis on the characterization of  
a conserved quantity in the large-scale limit.
Section III is devoted to the general solution of the large-scale  
perturbation dynamics while section IV presents the detailed mode  
structure for a variety of cosmologically relevant equations of  
state. A short summary is given in section V.

\section{Basic relations of a covariant perturbation theory}
We consider the cosmic medium characterized by an energy-momentum tensor  
with perfect fluid structure,
\begin{equation}
T _{mn} = \rho u _{m}u _{n} + p h _{mn}\ ,
\mbox{\ \ \ \ }
\left(m,n... = 0,1,2,3 \right)\ ,
\label{1}
\end{equation}
where $\rho $ is the energy density, $p$ is the pressure, $u_{m}$  
is the four-velocity ($u ^{m}u _{m} = -1$), and
$h_{mn} = g_{mn} + u_{m}u_{n}$ the spatial projector on surfaces
orthogonal to the 4-velocity.
We recall that the structure (\ref{1}) is also valid for a  
minimally coupled scalar field with the identifications
\begin{equation}
\rho  = \frac{1}{2}\dot{\phi }^{2} + V \left(\phi  \right)\ ,
\mbox{\ \ \ \ }
p  = \frac{1}{2}\dot{\phi }^{2} - V \left(\phi  \right)\ ,
\label{2}
\end{equation}
and
\begin{equation}
u _{i} =  - \frac{\phi _{,i}}
{\sqrt{-g ^{ab}\phi _{,a}\phi _{,b}}}\ ,
\label{2a}
\end{equation}
where $\dot{\phi } \equiv  \phi _{,a}u ^{a} = \sqrt{-g ^{ab}\phi  
_{,a}\phi _{,b}}$ and $V \left(\phi  \right)$ is the scalar field  
potential.

Local energy momentum conservation $T ^{ik}_{\ ; k} = 0$, implies
\begin{equation}
\dot{\rho } = - \Theta\left(\rho + p\right) \ ,
\mbox{\ \ \ \ }
\left(\rho  + p\right)\dot{u}_{}^{m} = 
- p_{,k}h^{mk}\ .
\label{3}
\end{equation}
Here,  
$\Theta \equiv u^{i}_{\ ;i}$
is the fluid expansion and 
$\dot{u}^{m} \equiv u^{m}_{;n}u^{n}$  the fluid acceleration.

Additionally, we make use of the Raychaudhuri equation
\begin{equation}
\dot{\Theta} + \frac{1}{3}\Theta^{2} 
+ 2\left(\sigma^{2} - \omega^{2}\right) - \dot{u}^{a}_{;a} 
- \Lambda 
+ \frac{\kappa}{2}\left(\rho  + 3p\right) 
= 0 \ ,
\label{4}
\end{equation}
where $\Lambda $ is the cosmological constant and
$\kappa$ is Einstein's gravitational constant.
The magnitudes of shear and vorticity are defined by
\begin{equation}
\sigma^{2} \equiv \frac{1}{2}\sigma_{ab}\sigma^{ab}\ ,\ \ \ \ \
\omega^{2} \equiv \frac{1}{2}\omega_{ab}\omega^{ab}\ ,
\label{5}
\end{equation}
with
\begin{equation}
\sigma_{ab} = h_{a}^{c}h_{b}^{d}u_{\left(c;d\right)} 
- \frac{1}{3}\Theta_{}h_{ ab}\ ,\ \ \ \ \
\omega_{ab} = h_{a}^{c}h_{b}^{d}u_{\left[c;d\right]} 
\ .
\label{6}
\end{equation}
The 3-curvature scalar of the projected metric,
\begin{equation}
{\cal R} = 2 \left(- \frac{1}{3}\Theta ^{2}
+ \sigma ^{2} - \omega ^{2} + \kappa \rho  + \Lambda \right)\ ,
\label{7}
\end{equation}
reduces to the 3-curvature of the surfaces orhogonal to $u ^{a}$ for   
$\omega = 0$.
The homogeneous and isotropic Friedmann-Lema\^{\i}tre-Robertson-Walker 
(FLRW) universes correspond to the special case
$\sigma = \omega = \dot{u}^{a} = 0$.

Suitable covariant variables to characterize spatial inhomogeneities are 
\cite{EB,Jack,ZCQG,ZPRD}
\begin{equation}
D _{a} \equiv \frac{a h ^{c}_{a}\rho _{,c}}{\rho + p} \ ,
\mbox{\ \ \ \ }
P _{a} \equiv \frac{a h ^{c}_{a}p _{,c}}{\rho + p} \ ,
\mbox{\ \ \ \ }
t _{a} \equiv a h ^{c}_{a} \Theta _{,c} \ ,
\label{8}
\end{equation}
where $a$ is a length scale generally defined by $\Theta \equiv  3  
\dot{a}/a$.
The quantities $D _{a}$ and $P _{a}$ represent fractional, comoving  
(with the fluid four-velocity) energy density and pressure  
perturbations, respectively.
Inhomogeneities in the expansion are described by the quantity $t _{a}$. 

In a homogeneous and isotropic FLRW universe (superscript 0) the  
Friedmann equation
\begin{equation}
\kappa \rho ^{^{\left(0 \right)}}
= \frac{1}{3}\left(\Theta ^{^{\left(0 \right)}} \right)^{2}
+ \frac{1}{2}{\cal R}^{^{\left(0 \right)}} - \Lambda \ ,
\label{9}
\end{equation}
holds and the  Raychaudhuri equation (\ref{4}) reduces to
\begin{equation}
\dot{\Theta }^{^{\left(0 \right)}}
+ \frac{3}{2}\kappa \left(\rho ^{^{\left(0 \right)}}
+ p ^{^{\left(0 \right)}}\right)
= \frac{1}{2} {\cal R}^{^{\left(0 \right)}} \ .
\label{10}
\end{equation}
The three-curvature in this case is known to be
${\cal R}^{^{\left(0 \right)}} = 6 k/a ^{2}$,
where $k = 0,\pm 1$ and $a$ now coincides with the scale factor of  
the Robertson-Walker metric.

In order to describe the dynamics of the inhomogeneities in terms of 
$D _{a}$, $P _{a}$ and $t _{a}$ we differentiate the first Eq.  
(\ref{3}),
project orthogonal to $u _{a}$ and multiply by $a$.
The left-hand side of the resulting equation may be written as (cf.  
\cite{Jack})
\begin{equation}
a h ^{c}_{m}\dot{\rho }_{,c} = h ^{a}_{m}
\left(a h _{a}^{c}\rho  _{,c} \right)^{\displaystyle \cdot}
- a \dot{u}_{m}\dot{\rho}
+ \left(\omega ^{c}_{\ m} + \sigma ^{c}_{\ m} \right)
a h ^{n}_{c}\rho  _{,n}\ ,
\label{10a}
\end{equation}
where we have used the well-known decomposition
\[
u _{i;n} = - \dot{u}_{i}u _{n} + \sigma _{in} +  \omega _{in}
+ \frac{\Theta }{3}h _{in}
\]
of the covariant derivative of the four-velocity.

Restricting ourselves to first-order deviations from homogeneity  
and isotropy
we find \cite{ZCQG}
\begin{equation}
\dot{D}_{\mu }
+ \frac{\dot{p}}
{\rho  + p} D _{\mu }
+ t _{\mu } = 0 \ ,
\mbox{\ \ \ \ \ }
\left(\mu ,\nu... = 1,2,3 \right)
\ .
\label{11}
\end{equation}
Analogously, one obtains from Eq. (\ref{4}),
\begin{equation}
\dot{t}_{\mu }  =
- \frac{2}{3} \Theta t _{\mu }
- \frac{\kappa }{2} \left(\rho
+ p \right) D _{\mu }
- \left(\frac{1}{2}{\cal R} +
\frac{\nabla ^{2}}{a ^{2}}  \right) P _{\mu }  \ .
\label{12}
\end{equation}
Combining the last two equations and
defining the sound velocity $c _{s}$ in a standard way as
$c _{s}^{2} = \dot{p}/\dot{\rho }$, the first-order inhomogeneities  
are governed by the second-order equation
\begin{eqnarray}
&&\ddot{D}_{\mu } + \left(\frac{2}{3} - c _{s}^{2}\right)
\Theta \dot{D}_{\mu }
- \left[\left(c_{s}^{2} \right)^{\displaystyle \cdot}\Theta 
\right.\nonumber\\
&&+ \left. \left(\frac{\kappa }{2}\left(\rho - 3 p\right) + 2 \Lambda\right)
c _{s}^{2} + \frac{\kappa }{2} \left(\rho + p\right) \right] D _{\mu } 
=  \frac{\nabla ^{2}}{a ^{2}} P _{\mu }\ .
\label{13}
\end{eqnarray}
For a fluid (superscript f) one has
$P _{a}^{^{\left(f \right)}} = c_{s}^{2} D _{a}^{^{\left(f \right)}}$
and Eq. (\ref{13}) corresponds to Jackson's \cite{Jack} equation (57). 
For a scalar field
(superscript s),  because of $h ^{c}_{a}\phi _{,c} = 0$, the  
potential term neither contributes to $D _{a}^{^{\left(s \right)}}$  
nor to
$P _{a}^{^{\left(s \right)}}$ and, consequently,
$P _{a}^{^{\left(s \right)}} =  D _{a}^{^{\left(s \right)}}$
is valid \cite{ZPRD} which, if used in Eq. (\ref{13}), results in a  
closed equation for
$D _{\mu }$ as well.

The well-known second-order differential equation (\ref{13})  
together with a corresponding expression for $P _{\mu }$ provides us  
with a complete description of linear scalar perturbations in  
perfect fluid and scalar field universes.
Our aim here is to find the general solution of equation (\ref{13})  
in the large-scale limit for spatially flat $\left(k = 0 \right)$  
cosmological models.
To this purpose we rewrite the perturbation dynamics in terms of a  
different basic variable.

\section{Solving the large-scale perturbation dynamics}
As follows from the definitions (\ref{8}), the quantities $D _{a}$,  
$P _{a}$
and $t _{a}$ are defined with respect to comoving hypersurfaces.
It was shown in \cite{ZCQG} that the perturbation dynamical  
description simplifies if written in terms of covariant variables  
defined with respect to hypersurfaces of constant curvature,  
constant expansion, or constant energy density.
These variables correspond to certain combinations of  
Ellis-Bruni-Jackson type variables.
For example, the exactly defined covariant quantity
\begin{equation}
D _{a}^{\left(ce \right)} \equiv
D _{a} - \frac{\dot{\rho }}{\rho + p}
\frac{t _{a}}{\dot{\Theta }}
\label{14}
\end{equation}
represents in first order the fractional, spatial gradient of the  
energy density on hypersurfaces of constant expansion (superscript  
ce)\cite{ZCQG,ZPRD}.
Rewriting the linear perturbation dynamics in terms of
$D _{a}^{\left(ce \right)}$ and restricting ourselves to $k = 0$,  
one obtains
\begin{equation}
\left[a ^{2}\dot{\Theta }
D _{\mu }^{\left(ce \right)}\right]^{\displaystyle \cdot}
=   - a ^{2}\Theta \frac{\nabla ^{2}}{a ^{2}}P _{\mu } \ .
\label{15}
\end{equation}
Except for the spatial pressure gradient on the right-hand side of
Eq. (\ref{15}) all terms of the dynamical perturbation equation may  
be included into a first time derivative.
With $D _{\mu } = D _{\left(m \right)}\nabla _{\mu }
Q _{\left(m \right)}$,
$P _{\mu}
= P _{\left(m \right)}
\nabla _{\mu }Q _{\left(m \right)}$ and
$D _{\mu }^{\left(ce\right)}
= D _{\left(m \right)}^{\left(ce\right)} \nabla _{\mu }Q _{\left(m  
\right)}$
(and corresponding relations for the other perturbation quantities),  
where
the $Q _{\left(m \right)}$ satisfy the Helmholtz equation $\nabla ^{2}Q
_{\left(m \right)} = - m ^{2} Q _{\left(m \right)}$, the quantity  
$m$ is related to the physical
wavelength by $\lambda = 2 \pi a/m$
(\cite
{KoSa,DB,LiKha,Harr}).
It follows that the spatial gradient terms  on
the right-hand side of Eq.(\ref{15})
may be neglected on large perturbation scales ($m \ll 1$)
and the quantity
$a ^{2}\dot{\Theta }
D _{\left(m\right) }^{\left(ce \right)}$ is a conserved quantity both   
for a perfect fluid and a scalar field.

Denoting the conserved quantity by $-E _{\left(m \right)}$, i.e.
\begin{equation}
a ^{2}\dot{\Theta }D _{\left(m \right) }^{\left(ce \right)}
\equiv   - E _{\left(m \right)}  = const \ ,
\mbox{\ \ \ \ }
\left(m \ll 1 \right)\ ,
\label{16}
\end{equation}
and taking into account the relation
\begin{equation}
\frac{\dot{\Theta}}{\Theta }D _{\mu }^{\left(ce \right)}
= - \dot{D}_{\mu }
+  \left[ \frac{\dot{\Theta }}{\Theta } +
c _{s}^{2}\Theta  \right]
D _{\mu }\
\label{17}
\end{equation}
between $D _{\mu }^{\left(ce \right)}$ and $D _{\mu }$ which  
follows from the definition (\ref{14}) of $D _{\mu }^{\left(ce  
\right)}$ and Eq. (\ref{11}),
the equation to solve is
\begin{equation}
\dot{D}_{\left(m \right) } - \left(\frac{\dot{\Theta }}{\Theta }
+ c _{s}^{2} \Theta  \right)D _{\left(m \right)}
= \frac{E _{\left(m \right) }}{a ^{2}\Theta }\ ,
\mbox{\ \ \ \ \ }
\left(m \ll 1 \right)\ .
\label{18}
\end{equation}
With $\Theta = 3 \dot{a}/a$ and the abbreviation
\begin{equation}
q \equiv  \int_{}^{} \left(c _{s}^{2} \right)^{\displaystyle \cdot}
\ln a ^{3} dt
\label{19}
\end{equation}
the general solution to Eq. (\ref{18}) for $m \ll 1$ is
\begin{eqnarray}
D _{\left(m \right) } &=& \Theta a ^{3 c _{s}^{2}} \exp{\left\{- q  
\right\}}\cdot
\nonumber\\
&&\cdot\left[\int^{t} dt \left(
\frac{E _{\left(m \right) }}{\Theta ^{2}a ^{2 + 3 c _{s}^{2}}} \right)
\exp{\left\{q \right\}} + C _{\left(m \right)}\right]\ ,
\label{20}
\end{eqnarray}
where $C _{\left(m \right)}$ is an integration constant.
For many purposes it is sufficient to consider
$c _{s}^{2} \approx const$ in which case $q = 0$ and the solution  
(\ref{20}) reduces to
\begin{equation}
D _{\left(m \right)} = \Theta a ^{3 c _{s}^{2}}
\left[\int^{t} dt \left(
\frac{E _{\left(m \right)}}{\Theta ^{2}a ^{2 + 3 c _{s}^{2}}} \right)
+ C _{\left(m \right)}\right]\ , 
\mbox{\ \ }
\left(m \ll 1 \right) .
\label{21}
\end{equation}
Formula (\ref{20}) comprises the entire large-scale linear  
perturbation dynamics for arbitrary equations of state.
It is the main result of this paper.
In the following section we use the expression (\ref{20}) (or the  
special case (\ref{21})) to derive the mode structure for different  
cosmologically relevant equations of state.

\section{Cosmological mode structure in the large-scale limit}
\subsection{Vanishing cosmological constant ($\Lambda = 0$)}
Let us assume equations of state $p = \left(\gamma - 1 \right)\rho  
$ with  constant values of $\gamma $ in the range
$0 \leq \gamma \leq 2$, i.e. 
$c _{s}^{2} \equiv  \dot{p}/\dot{\rho } = \gamma - 1$.
The case $\gamma = 0$, equivalent to $p = - \rho $ and
$c _{s}^{2} = - 1$, is characterized by $\rho  = const$,
$\Theta \equiv  3H = const$, i.e. $a \propto \exp{\left[Ht \right]}$. 
Use of these dependences in formula (\ref{21}) yields a behaviour
$\propto a ^{-2}$ and $\propto a ^{-3}$ for the cosmological modes.  
Any perturbation about the de Sitter spacetime is exponentially  
damped \cite{ZPRD}.

For $0 < \gamma \leq 2$ we have
$\rho \propto a ^{-3 \gamma }$ and, via Eq. (\ref{9}),
$a \propto t ^{2/\left(3 \gamma  \right)}$.
Since $\Theta \propto t ^{-1}$
one obtains a behaviour
$\propto a ^{3 \gamma - 2} \propto t ^{2 - 4/\left(3 \gamma   
\right)}$ and
$\propto a ^{3 \gamma /2 -3} \propto t ^{1 - 2/\gamma }$ for the  
large-scale modes \cite{Jack} which coincides with corresponding results in \cite{HH}. 

It follows that in the range $0 < \gamma  < 2/3$ where $a \propto t  
^{n}$ with $n > 1$ (power-law inflation) both perturbation modes  
are decaying.
The case $\gamma = 2/3$, corresponding to so-called K-matter
\cite{KOLB}, is  
characterized by $a \propto t$. The dominant mode turns out to be  
constant in this case \cite{APZ}
while the second mode decays as $a ^{-2} \propto t ^{-2}$.

In the interval $2/3 < \gamma < 2$ there is one decaying and one  
growing mode.
For $\gamma = 1$ (dust) one reproduces the well-known behaviour
$\propto a \propto t ^{2/3}$ and $\propto a ^{-3/2}\propto t ^{-1}$  
for the large-scale perturbation modes, while one obtains modes
$\propto a ^{2} \propto t$ and $\propto a ^{-1} \propto t ^{-1/2}$ for 
$\gamma = 4/3$ (radiation).

For $\gamma = 2$ (stiff matter) there is no decaying mode. The  
first term in (\ref{21}) grows as $ a ^{4}\propto t ^{4/3}$ while  
the second one remains constant.

None of these results is new. Our point here is to demonstrate the  
possibility of a unified representation of the cosmological mode  
structure via
formulae (\ref{20}) or (\ref{21}).
We believe this representation to be useful also under more general  
circumstances, where, e.g. $c _{s}^{2} $ varies in time or in  
multifluid cosmological models to be studied elsewhere.

\subsection{$\Lambda $-dominated universe}

For $\Lambda \gg \kappa \rho $ in Eq. (\ref{9}) we have
$\Theta = 3 H = const$, i.e. $a \propto \exp{\left[H t \right]}$.
Matter on this background is again characterized by
$p = \left(\gamma - 1 \right)\rho $ with
$c _{s}^{2} = \left(\gamma - 1 \right)$.
>From the general formula (\ref{21}) we find perturbation modes   
$\propto \exp{\left[-2Ht \right]}$ and
$\propto \exp{\left[3 \left(\gamma - 1 \right) \right]}$.
While the first mode exhibits the same behaviour (exponential  
damping) as the dominating mode for $\Lambda = 0$ and $\gamma = 0$,  
the behaviour of the second
mode strongly depends on $\gamma $.
It decays exponentially for any $\gamma < 1$, e.g. for K-matter  
\cite{APZ} with
$\gamma = 2/3$, but grows  for $\gamma > 1$, e.g. for radiation  
with $\gamma = 4/3$.
Comoving perturbations in relativistic matter in a $\Lambda  
$-dominated universe  are exponentially unstable.
A behaviour such as this was first found in \cite{ZMNRAS} by  
directly solving a second-order equation of the type (\ref{13}).
A $\Lambda $-dominated universe differs from the previous ones with  
$\Lambda = 0$ in so far as the background dynamics according to  
(\ref{9})  is completely determined by $\Lambda $ in the case  
considered here ( $\kappa \rho \ll \Lambda $).
On the other hand, differentiating Eq. (\ref{4}) in order to obtain 
Eq. (\ref{12}) the cosmological constant drops out and does not  
enter the perturbation dynamics  explicitly.
There are perturbations only in the component which is dynamically  
negligible in the background but not, by definition, in $\Lambda $  
which, however, determines the evolution of $a$ and $\Theta $ in  
formula (\ref{21}).
The mentioned instability refers to this kind of perturbations.

\section{Summary}
We have presented a compact, covariant version of linear  
cosmological perturbation theory for media characterized by an  
energy-momentum tensor of the perfect fluid type.
The general solution of the large-scale perturbation dynamics in a  
spatially flat universe was obtained as a closed time integral for  
comoving energy density perturbations, resulting in a comprehensive  
picture of the cosmic mode structure for different cosmological  
epochs including exponential  and power-law inflation as well as a  
K-matter period, the standard FLRW cases and matter perturbations in  
a $\Lambda $-dominated universe.

\acknowledgments
This paper was supported by the Deutsche Forschungsgemeinschaft,
the Spanish Ministry of Education under grant PB94-0718 and  
the NATO (grant CRG940598).

\end{document}